\documentclass[twocolumn,amsmath,amssymb,aps]{revtex4-2}

\usepackage{gensymb} 
\usepackage{graphicx}
\usepackage{float}
\usepackage{dcolumn}
\usepackage{bm}
\usepackage{hyperref}

\usepackage{enumitem} 

\usepackage{xcolor}

\begin{document}

\preprint{APS/123-QED}

\title{Experimental straintronics in nanotube quantum dots}

\author{L. Huang}
\author{I.~G. Rebollo}%
\author{A.~R. Champagne}%
 \email{a.champagne@concordia.ca}
\affiliation{%
 Department of Physics, Concordia University, Montr\'{e}al, Qu\'{e}bec, H4B 1R6, Canada
}%

\date{\today}%

\begin{abstract} %
Single-wall carbon nanotubes (SWCNTs) are narrow ribbons of graphene with atomically precise edges and a single quantum transport channel, at experimentally-relevant dopings. This makes them ideal systems to harness \textit{quantum transport straintronics} (QTS), i.e.\ using mechanical strain to control accurately quantum transport. We present QTS data from three single-wall carbon nanotube quantum dot (SWCNT-QD) transistors over a broad range of \textit{in-situ} tunable and reversible uniaxial strain ($\Delta\varepsilon_\text{mech}\approx$ 0 to 3\%). We first present the nanofabrication of the suspended SWCNT transistors whose channel lengths are $\approx$ 30 nm. The channels are strained by moving gold clamps holding firmly the nanotubes. We present detailed charge transport data, $dI/dV_{\text{B}} - V_{\text{B}} - V_{\text{G}}$ and $dI/dV_{\text{B}} - V_{\text{B}} - \Delta\varepsilon_\text{mech}$, showing a large mechanical-gating effect of the SWCNT-QDs. The precise reversibility of the data, and their agreement with QTS theory, confirms that the tubes are strained elastically. We demonstrate that the mechanical control of the QD doping is not due to capacitive-gating effects, but to \textit{quantitatively predictable} bandstructure changes including a strain-tunable bandgap. This precise mechanical control of the doping and bandgap of SWCNT-QDs could find applications in qubits, condensed matter physics, and homojunction molecular transistors.
\end{abstract}

\maketitle

Experimental control of quantum transport, and its quantitative agreement with theory, in one- and two-dimensional materials (1DMs/2DMs) requires very low and uniform disorder across entire devices\cite{Rhodes19,CastroNeto09}. Such a level of mastery has been achieved for charge-dopant disorder \cite{Banszerus16}, but not for mechanical (strain) disorder\cite{Kazmierczak21}, in most 2DMs and their nanotubes/nanoribbons. Quantum transport straintronics (QTS) aims to both remove the disruptive effects of disordered (uncontrolled) strain fields on quantum transport, as well as to add precisely engineered strain fields in 2DMs/1DMs to tune their properties (band gap, charge doping, gauge potentials, symmetry breaking) for quantum electronics applications \cite{Kim26,Huang25,Boland24,McRae24,Hou24,Miao21,McRae19,Amorim16, Guinea10}. Ideal quantum straintronics transistors would permit a complete control of the magnitude and phase of their quantum current both mechanically and electrostatically, and have \textit{in-situ} tunable lattice symmetry breaking. Technological applications of QTS in 1D nanotubes/nanoribbons, graphene, transition metal dichalcogenides (TMDs), and twisted 2DMs would include optimizing qubits and quantum circuits \cite{Riechert25,Chen23,Alfieri23, Tormo_Queralt22, Mergenthaler21_2, Banszerus21, Khivrich20, Liu19}, molecular transistors \cite{Wang25,van_der_Poel24,Evers20}, spintronics \cite{Pal23, Li20, Wu18, Hanakata18, Molle17, Kuemmeth08}, strain-tunable superconductivity\cite{Hou24,Kapfer23,Khanjani18}, and topological switches \cite{Kim23, Zhang23, Du21, Moulsdale20, Mutch19,Efroni17}.

There have been a few recent demonstrations of well-controlled QTS in graphene transistors \cite{McRae24, Wang21}. However, overall the experimental validation of many exciting QTS predictions \cite{Kim23,Miao21} has been a major challenge. Two significant limitations to exploring \textit{quantitative} QTS are that 2DM devices usually have atomically disordered edges which scramble the phase of their charge carriers, and the mesoscopic widths of most 2DM devices mean that many subbands (transverse momentum modes) contribute to their charge transport. Because the quantum phase of each mode is impacted differently by mechanical strain \cite{McRae19,Fogler08}, many subbands make a precise control of QTS complex. Single-wall carbon nanotubes (SWCNTs) \cite{Laird15} and other nanotubes \cite{An24,Arenal07} naturally resolve both of these issues as they have perfect transverse (periodic) boundary conditions, and their very narrow width (circumference) leads to a single transport subband being available at experimentally relevant charge dopings. Until now, only the most basic and room-temperature strain-dependent properties of SWCNTs, such as bandgaps, have been studied experimentally \cite{Minot03, Huang08} despite that they are widely used in industry \cite{Goerzen26}. Thus, SWCNTs represent a major unexplored opportunity for QTS experiments. For instance, recent theoretical work \cite{Huang25} foresees that transistors made with a SWCNT will permit a full \textit{mechanical} control of the quantum phase of their coherent charge current. Previous work on SWCNTs also suggested a broad strain tunability of their electron-phonon coupling \cite{Mariani09} and spin-orbit coupling \cite{Kuemmeth08}.

In addition to being a rolled-up 2DM, SWCNTs are also single-molecules which can naturally host quantum dots acting as zero-dimensional (0D) systems. The long length of nanotubes allows a precise alignment of the tubes, and permits to use the same tube for the contacts (source/drain) and channel of the device. This enables highly reproducible fabrication of QDs which nearly match the size and energy scales of traditional single-molecule transistors \cite{van_der_Poel24}. It is expected that straining a firmly clamped SWCNT-QD can be used to mechanically control its charge state \cite{Huang25}. This approach would permit to tune the doping of single-molecule transistors \cite{Evers20} even when their short channel is electrostatically shielded by their environment.
\begin{figure*}
\includegraphics[scale=1]{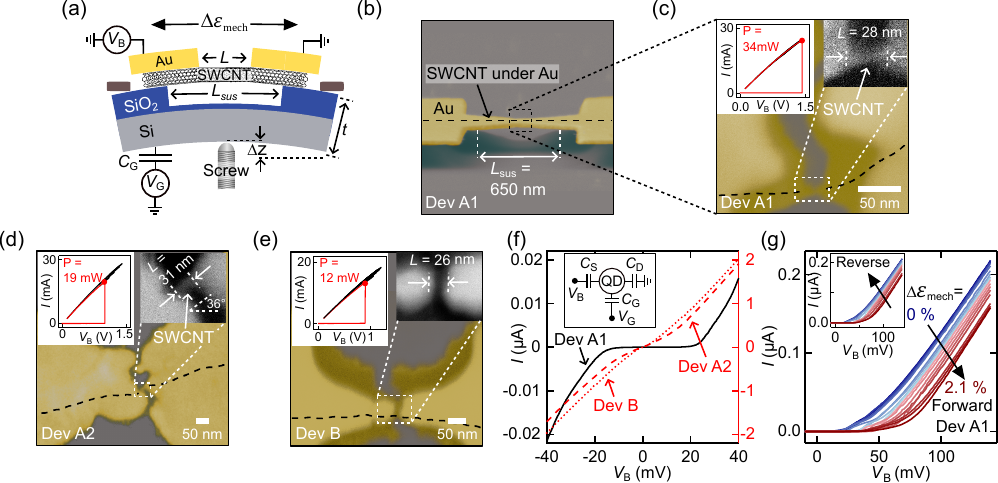}
\caption []{Instrumentation and SWCNT-QD transistors for quantum transport straintronics (QTS). (a) Side-view schematic of our instrumentation loaded in a low-temperature cryostat. A push screw bends slightly a very thin Si substrate, applying an \textit{in-situ} mechanical strain $\Delta\varepsilon_{\text{mech}}$ to the suspended SWCNT channel. A DC circuit is used for transport measurements. (b) A tilted-SEM image (80°) of the suspended SWCNT-under-gold breakjunction in Device A1 (before electromigration). The suspension length is $L_{\text{sus}}=$ 650 nm. (c) Zoom-in view of the breakjunction in (b) after electromigration. The top left inset shows the electromigration process for Device A1, and the top right inset the small scale SEM image of the naked-SWCNT channel created during this process. (d)-(e) show the same information as in (c) for Devices A2 and B, respectively. (f) $I$-$V_\text{B}$ data at 4 Kelvin, after electromigration, in Devices A1 (black, solid), A2 (red, dashed), and B (red, dotted). The inset is a schematic of our QDs' electronics circuit. (g) $I$-$V_\text{B}$ data in Device A2 versus increasing $\Delta\varepsilon_{\text{mech}}$. The inset shows the reverse mechanical sweep data, demonstrating hysteresis-free reversibility.}
\label{Fig1}
\end{figure*}

In this work, we report QTS data from three SWCNT-QD transistors over a broad range of \textit{in-situ} tunable and reversible uniaxial strain ($\Delta\varepsilon_\text{mech}\approx$ 0 to 3 \%). We first present the nanofabrication of the devices, whose channels are strained by moving gold clamps holding the nanotubes. We discuss detailed transport data showing that the charge states of the QDs are tunable via strain. The data are fully reversible and reproducible. We then analyze and model the data with QTS theory \cite{Huang25}, and find that the mechanical-gating effect stems from predictable bandstructure changes described by scalar and vector potentials.

Figure \ref{Fig1} presents the instrumentation and suspended SWCNT transistors we designed and fabricated to study QTS in SWCNTs. The geometry of our platform is shown in Fig. \ref{Fig1}(a). The details of our instrumentation were reported previously \cite{McRae24}. In summary, it operates inside a low-temperature cryostat to precisely bend a Si substrate (thickness $t = $ 200 $\mu$m) and applies a uniaxial mechanical strain $\Delta\varepsilon_\text{mech}$ to our suspended SWCNT channels of length $L\approx$ 30 nm, and diameter $d\approx$ 2 nm. The nanotubes are held mechanically by gold clamps (50 nm-thick films). We have shown previously, in similar suspended SWCNT transistors (without tunable strain) \cite{McRae17}, that the source and drain electrodes for the charge carriers are sections of the same nanotube located under the gold clamps. This creates a complete source-channel-drain homojunction transistor from a single SWCNT. The source and drain contacts have a charge doping set via charge transfer from the gold film \cite{McRae17, Chaves14, Heinze02}. The overlap length between the gold clamps and each nanotube contact is a few $\mu$m-long and provides slippage-free clamping, as we will demonstrate below. Details about the micro and nanofabrication of the SWCNT devices can be found in Supporting Information S1 \cite{SI}.

We studied DC transport in these transistors at temperature $T=$ 4 Kelvin. We applied a bias voltage, $V_{\text{B}}$, to the source electrodes and measured the current, $I$, at the drain. The channel's Fermi energy (doping) was controlled via a gate voltage, $V_{\text{G}}$, applied to the Si backplane. A unique feature of our device design is that the gold clamps are also suspended. These clamps acted as cantilever arms which amplified the mechanical strain imparted to the SWCNT channel upon bending the substrate. The total suspension length of our transistors was $L_{\text{sus}}\approx$ 700 nm (Fig. \ref{Fig1}(a)). The strain applied to the channel is given by $\varepsilon_{\text{mech}}= \Delta x/L= (3L_{\text{sus}}t/D^2)\Delta z/L$, where $\Delta x$ is the uniaxial stretching of the tube, $D\approx$ 8.2 mm is the distance between the substrate anchoring points, and $\Delta z$ is the motion of the pushing screw bending the substrate. Over a large number of previous experiments, we found that a maximum $\Delta z_{\text{break}}\approx$ 260 $\mu$m could be applied before substrate failure \cite{McRae24}. In the present experiment, we applied up to $\Delta z=$ 100 $\mu$m and reached a maximum $\varepsilon_{\text{mech}}$ of $\approx$ 3 $\%$ in Device B which had a 26-nm-long channel and $L_\text{sus} \approx 890$ nm. It is worth mentioning that any gate-voltage induced strain is truly negligible due to the very short length of our channels \cite{McRae19}.

Figure \ref{Fig1}(b) shows a tilted (80$\degree$) scanning electron microscope (SEM) image of Device A1, before we used electromigration to create a naked SWCNT channel. The SWCNT, located under the gold film, is outlined with a dashed line. The measured $L_{\text{sus}}$ was 650 $\pm$ 45 nm. SEM images of the other two samples, Devices A2 and B, are shown in Supporting Information S1 and give $L_\text{sus}=$ 580 $\pm$ 55 nm and 890 $\pm$ 25 nm, respectively. The final step of sample fabrication was a low-temperature feedback-controlled electromigration \cite{Island11, McRae17}, which we used to create a nanogap in the suspended gold film. Figure \ref{Fig1}(c) shows Device A1 after the electromigration step. The top-left inset is a summary of the $I$ - $V_\text{B}$ electromigration data, where a relatively large current was used to create a Joule-induced electromigration of the gold. The electromigration power for Device A1 was $P=$ 34 mW. In the main panel, and the top right inset, of Fig.\ref{Fig1}(c) we see the gap formed in the gold film and the $\approx$ 28 nm SWCNT channel bridging the gap. The position of the nanotube beneath the gold film (dashed lines) was determined precisely by aligning SEM images of the SWCNT before the gold deposition and after the electromigration (Supporting Information S1).

Figures \ref{Fig1}(d)-(e) show the electromigration step\cite{Island11} and SEM images for Devices A2 and B, respectively. Devices A1 and A2 were fabricated on the same SWCNT at different locations. This is important since it guarantees that these two devices have the same nanotube chirality (rolling angle) and diameter. As shown in Supporting Information S1, the measured tube diameters for the devices were 2.1 $\pm$ 0.5 nm (Devices A1-A2) and 1.6 $\pm$ 0.4 nm (Device B).

Following electromigration, and before making the transport measurements, we used Joule-annealing (Supporting Information S1) to remove leftover nanofabrication residues on the SWCNT channels. Figure \ref{Fig1}(f) shows transport data $I - V_{\text{B}}$ in all three devices at $V_{\text{G}}$ and $\Delta\varepsilon_{\text{mech}} =$ 0. The resistance of Devices A2 and B were $\approx$ 20 k$\Omega$, consistent with open QD behavior \cite{Liang01,McRae17,Lotfizadeh21} (i.e. semi-transparent contact-channel tunnel barriers). On the other hand, Device A1 was a closed QD and showed a clear Coulomb blockade region \cite{Laird15} (opaque tunnel barriers). The inset of Figure \ref{Fig1}(f) shows a lump-element electronics diagram for our QD devices. The electrostatic potential on the dot \cite{JarilloHerrero04} depends on the source, drain and gate capacitance, respectively labelled $C_{\text{S}}$, $C_{\text{D}}$, $C_{\text{G}}$.

Figure \ref{Fig1}(g) shows the $I - V_{\text{B}}$ transport data for Device A1 at various strain values $\Delta \varepsilon_{\text{mech}}$. We notice a very smooth decrease in the conductance of the device as a function of the increasing strain. The inset shows that the data for a reverse mechanical sweep (decreasing strain) precisely reproduces the forward data. This was true for all three devices, and the relevant data for Device A2 and B are shown in Supporting Information S1. This reproducibility was achieved over numerous forward and reverse mechanical sweeps. This is very strong evidence that the SWCNT channels were stretched elastically without any slippage of the tubes under the gold clamps. In the remainder of this work, we measured and calculated how mechanical strain predictably modulates quantum transport in these three SWCNT-QDs.

\begin{figure}
\includegraphics[scale=1]{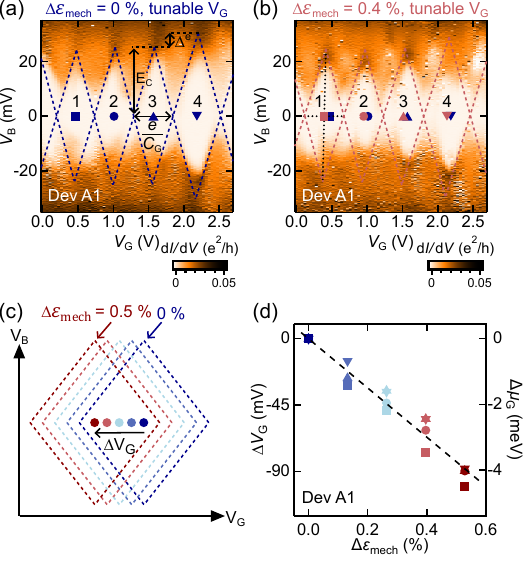}
\caption[]{Mechanically-tunable Coulomb diamond $\Delta V_{\text{G}}$ shifts. (a)-(b) $dI/dV_{\text{B}}-V_{\text{B}}-V_{\text{G}}$ data from Device A1 at $\Delta \varepsilon_{\text{mech}} =$ 0 and 0.4 \%, respectively.  (c) Outline of the Coulomb diamond 2 extracted from (a)-(b) and  similar data at $\Delta \varepsilon_{\text{mech}}$ ranging from 0 (blue) and 0.5 \% (red). The centers of the diamonds are labelled with circle markers. (d) The $\Delta V_\text{G}$ shift of diamonds 1 (squares), 2 (circles), 3 (triangles), 4 (downward triangles) versus $\Delta \varepsilon_{\text{mech}}$. The right-side axis shows the corresponding Fermi energy shift, $\Delta \mu_{G} = \alpha e \Delta V_\text{G}$. The dashed line is a linear fit.}
\label{Fig2}
\end{figure}

Figure \ref{Fig2} summarizes the electro-mechanical transport data measured in Device A1, and additional data for this device are in Supporting Information S2. Figure \ref{Fig2}(a) shows $dI/dV_{\text{B}}-V_{\text{B}}-V_{\text{G}}$ data at $\Delta \varepsilon_{\text{mech}} =$ 0. The methodology to determine the location of $\Delta \varepsilon_{\text{mech}} =$ 0 is presented in Supporting Information S2. The data show a clear series of Coulomb diamonds (dashed contours labelled $N=$ 1, 2, 3, 4). They correspond to QD charge ground states, where one electron is added to the QD as we move from a diamond to its neighbor on the right. We remark that the width and slopes of the diamonds give direct measurements of the capacitances $C_{S}$, $C_{D}$, and $C_{G}$ \cite{McRae17}. The widths of the diamonds 1-3 are $\approx$ 0.56 V, giving $C_\text{G} = $ 0.29 aF. Using a wire-over-plane electrostatic model for $C_\text{G}$, the extracted channel length is $L_\text{G} = 31$ nm. This is in good agreement with $L = 28$ nm obtained from Fig. \ref{Fig1}(c).

Figure \ref{Fig2}(b) shows a similar data set acquired at $\Delta \varepsilon_{\text{mech}} =$ 0.4 $\%$. The vertical and horizontal black dotted lines in diamond 1 illustrate how the center of a diamond was determined. We labeled the centers of diamonds 1, 2, 3, 4 with squares, circles, upward and downward triangles, respectively. The blue markers are for the $\Delta \varepsilon_{\text{mech}} =$ 0 diamonds and the light red markers for the $\Delta \varepsilon_{\text{mech}} =$ 0.4 $\%$ data. We observed that the center of each diamond shifted to the left with strain. This trend was confirmed by data sets at several other strain values, as shown in Supporting Information S2. Figure \ref{Fig2}c shows an enlarged version of the measured diamond shifts $\Delta V_{\text{G}}$ for diamond 2 as $\Delta \varepsilon_{\text{mech}}$ increased from 0 to 0.5 $\%$.

Figure \ref{Fig2}(d) summarizes all shifts of diamonds 1 (squares), 2 (circles), 3 (triangles), 4 (downward triangles) versus strain. We observe a linearly decreasing $\Delta V_{\text{G}}$ with $\Delta \varepsilon_{\text{mech}}$. The right hand-side axis gives the Fermi energy shift on the QD, $\Delta\mu_\text{G} = \alpha e \Delta V_\text{G}$, where $\alpha$ is the ratio between half-height of the diamond and its width. The dashed line is a linear fit to the data, and its slope gives $\Delta\mu_\text{G} = 7.5 \pm 0.3$ m$e$V/\%. As will be discussed below, this is consistent with predicted \cite{Huang25, Fogler08} strain-induced potentials $\phi_{\varepsilon}$ (workfunction shift) and $\mathbf{A_{\text{hop}}}$ (bandgap tuning) being added to the SWCNT-QD. Practically, this means that it is possible to gate (dope) the QD using purely mechanical means. One advantage of this ``mechanical-gating" is that unlike electrostatic-gating it cannot be screened by the QD environment, such as nearby metallic contacts, and could be useful in single-molecule electronics \cite{Evers20}.

\begin{figure}
\includegraphics[scale=1]{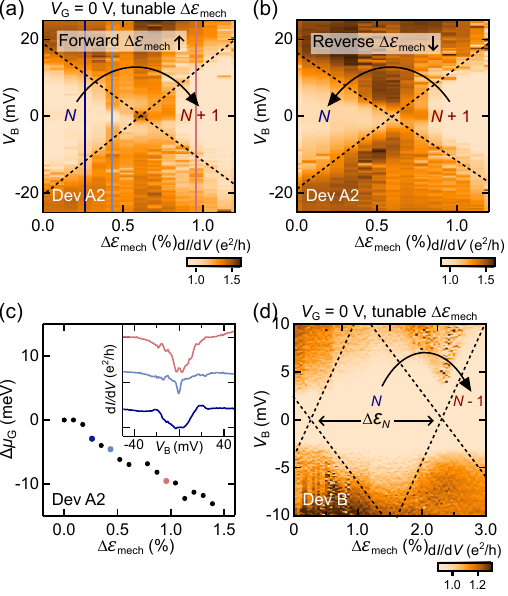}
\caption[]{Mechanical tuning of charge doping in SWCNT-QDs. (a)-(b) $dI/dV-V_\text{B}-\Delta\varepsilon_{\text{mech}}$ data from Device A2 for forward (increasing) and reverse (decreasing) $\Delta\varepsilon_{\text{mech}}$ sweeps, respectively. The labels $N$ and $N+1$ refer to the average number of electrons on the QD. (c) The main panel shows $\Delta \mu_{G}-\Delta\varepsilon_{\text{mech}}$ for Device A2. The inset shows $dI/dV-V_\text{B}$ data extracted from (a) at the location of the vertical cuts highlighted in dark blue, light blue, and red, respectively. (d) Same as (a), but for Device B.}
\label{Fig3}
\end{figure}

As shown in Fig. \ref{Fig3}, \textit{in-situ} strain control of the QD doping was observed even more dramatically in Devices A2 and B. It is important to note that although the $dI/dV\text{B} - V_{\text{B}} - \Delta \varepsilon_{\text{mech}}$ data in Fig. \ref{Fig3} may resemble the data in the color plots of Fig. \ref{Fig2}, the horizontal axis is no longer $V_{\text{G}}$ but rather $\Delta \varepsilon_{\text{mech}}$. These data were acquired at constant $V{_\text{G}}=$ 0. Fig. \ref{Fig3}(a) shows $dI/dV_{\text{B}} - V_{\text{B}} - \Delta \varepsilon_{\text{mech}}$ for Device A2 for a forward (increasing) strain sweep from 0 to 1.2 $\%$. Figure \ref{Fig3}(b) shows the reverse (decreasing) strain sweep. The data are highly reproducible and fully reversible, confirming again that there is no slippage of the tube under the gold clamps. Additional data and analysis details for Devices A2 and B are provided in Supporting Information S3.

We remark that the finite differential conductance $dI/dV$ near $V_\text{B} = 0$ indicates that Device A2 is an open quantum dot. Most importantly, well-defined diamond-like features are observed in the data, showing that we can use \textit{in-situ} tunable strain to add one full electron (on average) to the QD as we move the system from the regions labelled $N$ to $N+1$ in Figs. \ref{Fig3}(a)-(b).

The inset of Fig. \ref{Fig3}(c) shows the $dI/dV_{B}-V_{B}$ data traces extracted from the dark blue, light blue, and red vertical data cuts in Fig. \ref{Fig3}(a). As $\Delta\varepsilon_{\text{mech}}$ increases, the width of the low-conductance region shrinks and then grows again. A complete $dI/dV_{B}-V_{B}-V_{G}$ data set was recorded at each mechanical position (Supplemental Material S3). Using these data, we extracted the $V_{G}$-position of the charge degeneracy point (diamond crossing) for each $\Delta \varepsilon_{\text{mech}}$. As done above for Fig. \ref{Fig2}(d), we then calculated the channel's $\Delta\mu_\text{G} = \alpha e \Delta V_\text{G}$. Figure \ref{Fig3}(c) shows $\Delta\mu_\text{G}$ vs $\Delta \varepsilon_{\text{mech}}$ in Device A2. We observed that the Fermi energy of the QD could be tuned mechanically by over 13 meV. We emphasize that the ability to tune the charge state of a QD by a full electron is important in the context of quantum technologies \cite{Alfieri23}. For instance, adding a full electron to a closed QD (as in Fig. \ref{Fig2}) can modify its spin state.

Figure \ref{Fig3}(d) shows $dI/dV_{\text{B}} - V_{\text{B}} - \Delta \varepsilon_{\text{mech}}$ data for Device B. Once again, we see a clear strain control of the charge number on the QD. One striking difference is that increasing strain removes an electron (from $N$ to $N-$1), meaning that strain increases $\Delta\mu_\text{G}$. The sign of $\Delta\mu_\text{G}$ was observed directly in the $dI/dV_{B}-V_{B}-V_{G}$ data (Supporting Information S3) from the shift direction of the Coulomb diamonds in the $dI/dV - V_\text{B} - V_\text{G}$ maps.

\begin{figure*}
\includegraphics[scale=1]{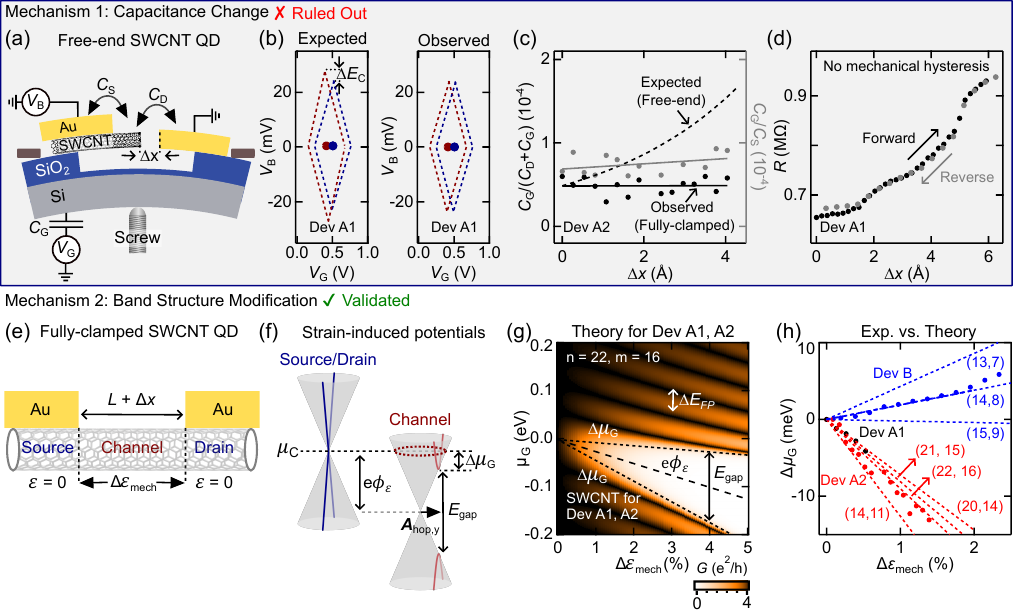}
\caption[]{Physical origin of mechanical gating in our SWCNT transistors. (a) Diagram showing the geometry of previously reported mechanically-controlled QDs (mechanical break-junctions). The QD was affixed to one metal contact (e.g. source) while free to move a distance $\Delta x$ away from another metal contact (e.g. drain). (b) The expected change, based on the geometry in (a), in the Coulomb diamond shape for Device A1 between $\Delta x =$ 0 ${\AA}$ (measured, blue) and $\Delta x =$ 4 ${\AA}$ (calculated, red). The right-hand side diamonds show that the actual measured diamonds at $\Delta x =$ 0 (blue) and 4 (red) ${\AA}$ did not change shape. (c) The measured slopes (solid markers) of the diamonds in Device A2 at various $\Delta x$. The solid lines are linear fits to the data. The dashed line shows a conservative estimate of the expected change in the diamond slopes based on the geometry in (a). (d) Measured $R$–$\Delta x$ data in Device A1 during forward and reverse sweeps show no mechanical hysteresis. (e) Diagram showing the geometry of the presently reported mechanically-controlled QDs. (f) Energy level and band structure of a SWCNT channel without (left cone) and with strain (right cone). A strain-induced scalar potential shifts the Dirac cone downwards by e$\phi_\varepsilon$. A strain-induced vector potential $A_{\text{hop,y}}$ creates a $k$-space shift leading to a bandgap $E_{\text{gap}}$. (g) Calculated $G-\Delta\varepsilon_\text{mech}-\mu_\text{G}$ for a SWCNT with a chirality (22,16) corresponding to Devices A1 and A2. (h) The measured $\Delta \mu_\text{G}$ vs $\Delta\varepsilon_\text{mech}$ in Devices A1, A2, and B. The dashed lines are theoretical calculations.}
\label{Fig4}
\end{figure*}

The observed mechanically-induced QD charge control is fundamentally different than in previous reports \cite{Cavena20,Parks07,Champagne05}, because it originates from band-structure modifications of the material and offers a predictable control. Figure 4 shows evidence of the stark difference between the previously reported mechanical-capacitance-tuning mechanism to gate QDs, Figs. \ref{Fig4}(a)-(d), and our strain-tunable band structure modification of QDs, Figs. \ref{Fig4}(e)-(h).

We briefly review the effect of mechanically-tunable QD capacitances on transport to understand why it cannot explain the data presented above. As shown in Fig. \ref{Fig4}(a), in previous mechanically-controlled QD (mechanical break-junctions) \cite{Cavena20,Parks07,Champagne05} the QD was affixed to one metal contact (e.g. source) and free to move a distance $\Delta x$ away from another metal contact (e.g. drain). Such a configuration leads to a mechanically-dependent QD-drain capacitance, $C_{\text{D}}$, and roughly constant QD-source, $C_{\text{S}}$, and QD-gate, $C_{\text{G}}$, capacitances. Based on this geometry (Supporting Information S4), the left-hand side of Fig. \ref{Fig4}(b) shows the expected change in the Coulomb diamonds in Device A1 between the measured data at $\Delta x =$ 0 $\AA$ (blue) and the expected (calculated) data for $\Delta x =$ 4 $\AA$ (red). We note that the heights and slopes of the expected red Coulomb diamond are significantly modified by $\Delta x$. The right-hand side of Fig. \ref{Fig4}(b) shows that the actual measured diamonds at $\Delta x =$ 0 (blue) and 4 $\AA$ (red) did not change shape, in clear disagreement with the capacitive model.

In Fig. \ref{Fig4}(c), we plot the measured slopes of the diamonds in Device A2 at various $\Delta x$. The upward slopes (black markers) and downward slopes (grey markers) correspond to the quantities $C_{\text{G}}/(C_{\text{D}}+C_{\text{G}})$ and $C_{\text{G}}/C_{\text{S}}$, respectively. The solid lines are linear fits to the data and show that the capacitances are nearly constant. The dashed line shows a conservative estimate (Supplemental Material S4) of the expected change in the diamond slopes if our QDs had the geometry shown in Fig. \ref{Fig4}(a). This further supports that the capacitances in our QDs do not change as we move the gold clamps.

This implies that both ends of our SWCNT-QDs are firmly anchored by the gold. Further evidence of this is visible in Fig. \ref{Fig4}(d) where we show the resistance, $R$, of Device A1 as we ramp up (forward) and down (reverse) the mechanical displacement $\Delta x$. The extremely good reversibility of the mechanical sweep is only compatible with a nanotube which is firmly clamped at both ends, and elastically stretched as a function of $\Delta x$. Collectively, the above data rule out capacitance changes as the primary mechanism for the mechanical dependence of our QDs' charge transport.

Figure \ref{Fig4}(e)-(h) show how strain-induced band structure modifications of the SWCNT-QDs lead to a quantitatively accurate description of our observed mechanical-gating effect. Figure \ref{Fig4}(e) shows how a uniform uniaxial strain $\Delta\varepsilon_{\text{mech}} = \Delta x / L$ is applied to the suspended nanotube channel \cite{McRae19}. Figure \ref{Fig4}(f) summarizes the expected strain-induced band structure modifications, as previously reported\cite{Huang25} and discussed in Supplemental Material S4. The solid blue (red) line in the left (right) Dirac cone of Fig. \ref{Fig4}(f) shows the lowest electronic subband in a quasi-metallic SWCNT (n$=22$, m$=16$) with $\Delta\varepsilon_{\text{mech}}=$ 0 ($\Delta\varepsilon_{\text{mech}}\neq$ 0). The Fermi level in the channel, $\mu_\text{C}$, is shown by the horizontal black dotted line.

When a uniaxial strain is applied to the SWCNT channel, the modifications to the transport subband (solid red line in Fig. \ref{Fig4}(f)) can be described by two new potentials. First, a scalar potential $\phi_{\varepsilon}$ shifts the entire dispersion (Dirac cones) down in energy relative to the unstrained source/drain SWCNT contacts. The magnitude of this scalar potential energy is \cite{Choi10,McRae24} $e\phi_{\varepsilon}= g_{\varepsilon}(1-\nu)\varepsilon_{\text{total}}$, where $\nu$ is the Poisson ratio 0.165 and $g_{\varepsilon}\approx 3.0$~eV, leading to a leftward shift of the Coulomb diamonds at a rate of $\approx$ 25 meV per 1 \% of mechanical strain. This effect is independent of the SWCNT chirality. We note that our transistor nanofabrication process has been shown to naturally selects metallic or very small bandgap nanotubes\cite{McRae17}, such as quasi-metallic ones. This is because metallic tubes have much higher contrast in scanning electron imaging (see Supporting Information S1), which we used to locate precisely our SWCNT channels before electron beam lithography.

Secondly, a vector potential $\boldsymbol{A_{\text{hop}}}$ shifts the $k$-position of the Dirac cones. If we define the $x$-axis along the length of the tube, then the $y$-component is $A_{\text{hop},y}=\frac{\beta\varepsilon(1+\nu)}{2a}\cos3\theta_h$, where $a=$1.42 $\text{Å}$ is the nearest-neighbor carbon-carbon distance without strain. $\theta_h$ is the angle between the tube's chiral vector $\boldsymbol{C}_{h}$ and the zigzag lattice direction (Supplemental Material S4) \cite{Huang25, Charlier07}. This $k$-shift of the subband leads to a strain tunable band gap $E_{\text{gap}}= 2\hbar v_{\text{F}}A_{\text{hop},y}$, as shown in Fig. \ref{Fig4}(f). This latter potential is chirality-dependent. The combination of the strain-induced scalar and vector potentials lead to a total Fermi level shift in the SWCNT channel given by $\Delta\mu_\text{G} = -e\phi_\varepsilon \pm E_\text{gap}/2$, where the sign of the second term depends on the type of charge carriers\cite{Huang25}.

To compare quantitatively the theoretical expectations with our measurements, Fig. \ref{Fig4}(g) shows a detailed calculation\cite{Huang25} of charge conductance $G = I/V_{\text{B}}$ (color scale) as a function of $\Delta\varepsilon_{\text{mech}}$ and $\mu_{\text{G}}$, for a quasi-metallic SWCNT with chirality (22,16). As we will see, this chirality is the likely candidate for Devices A1 and A2.  The calculated spectrum is very weakly dependent (Supplemental Material S4) on the minor chirality changes allowed to match both the measured tube diameter (2.1 nm) and the measured $\Delta\mu_{\text{G}} - \Delta\varepsilon_{\text{mech}}$ slope shown in Fig. \ref{Fig3}(c).

The dark stripes in Fig. \ref{Fig4}(g) indicate the energy-strain positions where the SWCNT-QD energy levels have resonant transmission. The central dashed line indicates the scalar potential energy shift, $e\phi_{\varepsilon}$. The white region corresponds to the strain-induced band gap resulting from $A_\text{hop,y}$. The sharp boundaries (dotted black lines) of the band gap region indicate the electrostatic potential energy shift $\Delta\mu_{\text{G}} - \Delta\varepsilon_{\text{mech}}$. The upper black dotted line corresponds to the conduction band edge shift (electrons), while the lower one corresponds to the valence band edge shift (holes).

Since the strain-induced channel's Fermi energy shift is chirality dependent, it can determine the chiral angle of a SWCNT. When combined with our tube diameter measurements, it constrains the possible tube chiral angles to just a few. Figure \ref{Fig4}(h) plots the experimental data for $\Delta\mu_\text{G}$ vs $\Delta\varepsilon_\text{mech}$ from Devices A1 (black), A2 (red) and B (blue). The experimental data were extracted as discussed previously for Figs. \ref{Fig2}(d) and \ref{Fig3}(c). Devices A1 and A2 show nearly identical negative linear trends, consistent with their fabrication on the same SWCNT. Device B, in contrast, shows a slope of different sign and magnitude.

By fitting the data (dashed lines), the chiral angles for the two SWCNTs are extracted. For Devices A1 and A2, we find $\theta =$ 25 $\pm$ 0.5 $\degree$, and for Device B, $\theta =$ 20.8 $\pm$ 0.3 $\degree$. Combining these precisely determined $\theta$ with the measured diameters of 2.1 $\pm$ 0.5 nm (Devices A1/A2) and 1.6 $\pm$ 0.4 nm (Device B), there are only two possible chiralities for Devices A1/A2 (22,16) or (21, 15) and one for Device B (14,8). Figure \ref{Fig4}(h) shows the quantitative agreement between the model (dashed lines) and the experimental data (markers) for all three devices.

In conclusion, single-wall carbon nanotube transistors have atomically precise transverse boundary conditions and a single transport subband, making them ideal systems to study and harness quantum transport straintronics. We presented transport data from three SWCNT-QD transistors, whose channel is $\approx$ 30-nm-long, over a broad range ($\approx$ 0 to 3 \%) of \textit{in-situ} tunable and reversible uniaxial strain.

The precise reversibility of the data presented confirms that the tubes are firmly clamped mechanically and strained elastically. Using $dI/dV_{\text{B}} - V_{\text{G}} - V_{\text{B}}$ data, we showed that the charge states of the QDs are widely tunable via strain. The agreement of the data with QTS theory calculations \cite{Huang25} confirmed that the mechanical tuning of the QDs stems from \textit{quantitatively} predictable band structure changes. These can be described as a strain-induced scalar potential which tunes the work function in the channel, and a vector potential which broadly adjusts the SWCNT's bandgap. This added mechanical-gating control of QD electronics could find applications in nanotube-based qubits \cite{Tormo_Queralt22} and quantum devices\cite{Mergenthaler21_2}, as well provide a reliable way to gate single-molecule electronics devices\cite{Evers20}.
\\
\\
\textbf{Data Availability} The data that support the findings of this study are included in the Figures of the main text and Supporting Information. They are also available from the corresponding author upon reasonable request.
\\
\\
\textbf{Author Contributions}
L.H. lead and did the majority of the fabrication of the devices, acquired most of the data presented, lead the data analysis and contributed to all other aspects of the work. I. G. R. contributed to the fabrication of the devices and to some of the data acquisition. A.R.C designed, supervised, and made significant contributions to all aspects of the work. A.R.C. wrote the manuscript with comments and inputs from all of the authors.
\\
\\
\textbf{Supporting Information}
The online version contains Supporting Information about the SWCNT device parameters, additional data from Devices A1, A2 and B, and additional discussion of data analysis and theoretical calculations.
\\
\\
\textbf{Acknowledgments}
This work was supported by NSERC (Canada), CFI (Canada), and Concordia University. We acknowledge usage of the LMF (Laboratoire de Microfabrication) at \'{E}cole Polytechnique Montr\'{e}al.
\\

\end{document}


\maketitle
\tableofcontents

\section{Additional information about the SWCNT devices}
\label{SM_Fig1}
In this section, we summarize key parameters of the SWCNT devices shown in Fig.~1(b)–(e) of the main text, and describe critical fabrication steps including the nanotube diameter measurement, gold clamp alignment, and junction suspension. We then describe the annealing process used to clean the SWCNTs. We note that all transport measurements reported in the main text were carried out after the final anneal of each device. Lastly, we present the mechanical calibration and additional data confirming reproducible and reliable mechanical motion using the instrumentation shown in Fig.~1(a) of the main text.

\subsection{Overview of the key parameters of our SWCNT devices}
We list the key parameters for the SWCNT devices presented in the main text in Table \ref{Table:SWCNT_parameters}. Devices A1 and A2 were fabricated on the same SWCNT, which is shown in Fig.\ \ref{Fig. S2}. The devices' channel lengths, obtained from SEM images, were around 30 nm for all devices. The suspension length $L_\text{sus}$ varied between 580 and 890 nm (see Fig. 1 in the main text and Fig.\ \ref{Fig. S2}). This variation arose from the width of the gold clamps (bow-tie junction before electromigration) which impacted the SiO$_2$ wet etching length underneath the devices.

For each of the three devices, the quantum dot capacitance ratios $C_\text{G}/C_\text{S}$ and $C_\text{G}/(C_\text{D}+C_\text{G})$ were extracted from the slopes of Coulomb diamonds in the $dI/dV$–$V_\text{G}$–$V_\text{B}$ maps (Figs. 2-3 in the main text, and Figs.\ \ref{Fig. S5}, \ref{Fig. S6}, \ref{Fig. S7}). Devices A2 and B showed significantly smaller capacitance ratios than Device A1, consistent with their weaker electrostatic gate coupling (due to screening from their electromigrated gold clamps).

\begin{table}[!htbp]
\centering
\includegraphics[scale=0.5]{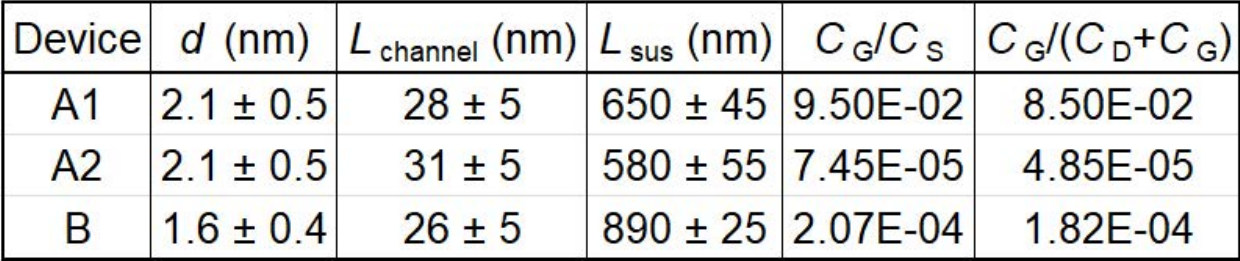}
\caption []{Key parameters of the SWCNT devices.}
\label{Table:SWCNT_parameters}
\end{table}

\subsection{Measurements of the SWCNT diameters}
\label{Subsec:SWCNT_Diameter}
\begin{figure}[!htbp]
\centering
\includegraphics[scale=1.2]{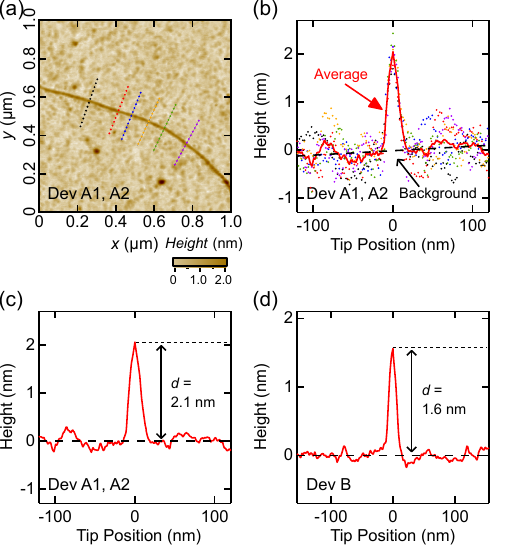}
\caption []{Measurements of the SWCNT diameters. (a) $1~\mu$m~$\times$~$1~\mu$m AFM image of the SWCNT in Devices A1 and A2. (b) Markers with different colors indicate data from different scan lines across the nanotube, as shown in (a). The red trace is the average of all data points, and the black dashed line is the linear fit of the substrate background. (c),(d) After removing the background tilt, the nanotube diameters were determined to be 2.1~$\pm$~0.5 and 1.6~$\pm$~0.4~nm for Devices A1/A2 and B, respectively.}
\label{Fig. S1}
\end{figure}

We measured the SWCNT diameters using AFM (atomic force microscopy). Figure~\ref{Fig. S1}(a) shows the height measurement over a $1~\mu$m~$\times$~$1~\mu$m area, where black, red, blue, gold, green, and purple dotted lines indicate the scan positions corresponding to the data in Fig.~\ref{Fig. S1}(b). To minimize the effect of substrate tilt, we averaged all data points (red curve) and fitted the regions away from the nanotube center to extract the background. After removing this background tilt, the nanotube diameter was obtained by measuring the height difference between the nanotube and the substrate as shown in Figs.~\ref{Fig. S1}(c)-(d). The nanotube for Devices~A1 and~A2 had a diameter of 2.1~$\pm$~0.5~nm, while that for Device~B had a diameter of 1.6~$\pm$~0.4~nm. The SWCNT diameters confirmed that the nanotubes were single-walled~\cite{Yamada06}.

\subsection{Alignment and suspension of the Au clamps}
\begin{figure}[!htbp]
\centering
\includegraphics[scale=1.2]{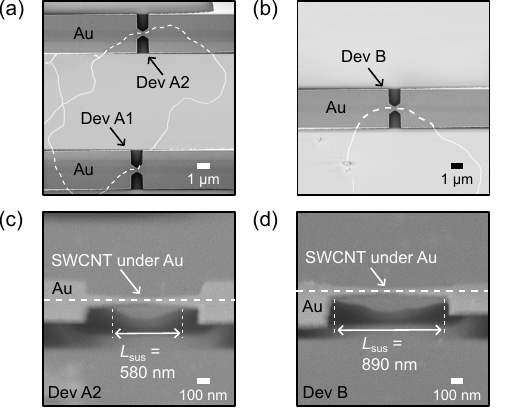}
\caption []{Gold clamps anchoring the suspended SWCNT channels. (a) Au clamps were aligned and deposited on the selected horizontal SWCNT sections for Devices A1 and A2. Dashed lines were drawn based on the superposition of SEM images of the SWCNT before and after Au clamp deposition. (b) Au clamps on top of the SWCNT section used for Device B. (c)-(d) Tilted (80$\degree$) SEM images of the suspended SWCNT-under-Au junctions in Devices A2 and B, respectively.}
\label{Fig. S2}
\end{figure}

Here, we provide additional details on the alignment and suspension of the Au clamps used in our suspended SWCNT quantum dots. Figure~\ref{Fig. S2}(a) shows the Au clamps deposited on the selected SWCNT sections for Devices~A1 and~A2. The nanotube direction is aligned with the chip bending direction (horizontal). The white dashed lines indicate the nanotube's position beneath the Au film, obtained by superimposing SEM images taken before and after the Au deposition. The Au clamps for Device~B are shown in Fig.~\ref{Fig. S2}(b). After removing the substrate oxide, the junctions were suspended as shown in Figs.~\ref{Fig. S2}(c)-(d) for Devices~A2 and~B, respectively. Device~A2 has a suspension length of 580~nm, whereas Device~B has a suspension length of 890~nm, which explains the larger mechanical strain range observed in Device B.

\subsection{Annealing of the SWCNT transistors}
\begin{figure}[!htbp]
\centering
\includegraphics[scale=1.2]{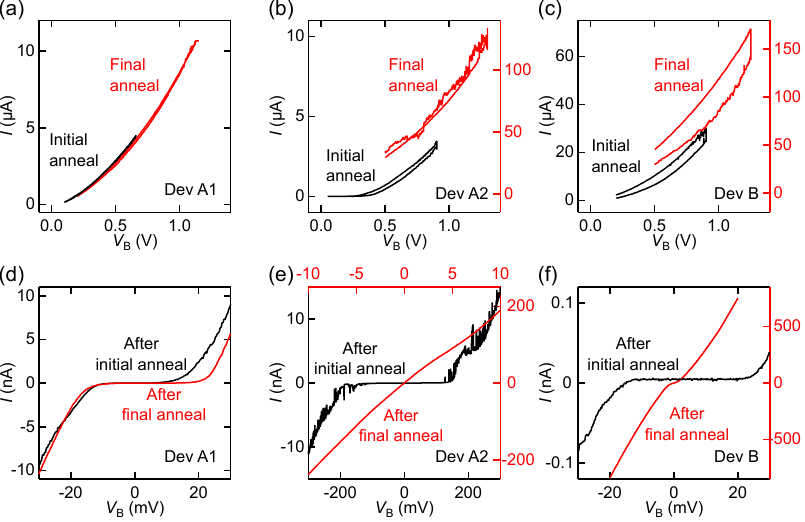}
\caption []{Joule annealing of the SWCNT transistors. (a)–(c) Annealing curves for Devices A1, A2, and B. (d)–(f) Black and red curves are bias sweeps after the initial and final anneal, respectively, for Devices A1, A2, and B. Devices A2 and B became highly conductive with minimal blockade, indicating open quantum dots, while Device A1 remained a closed quantum dot.}
\label{Fig. S3}
\end{figure}

After fabricating the suspended SWCNT-under-Au junctions, we loaded the devices into a cryostat and cooled them down to low-temperature. The SWCNT channels were then created via gold electromigration, as shown in Fig.~1 of the main text. Even after electromigration, residues from fabrication (such as resist) may remain on the SWCNT channels, degrading transport measurements. To remove these residues we used Joule annealing. We flowed a large current through the SWCNT, generating heat to raise the temperature in the suspended channels and anneal the contaminants. Figures~\ref{Fig. S3}(a)–(c) show the first and final annealing processes for each of Devices~A1, A2, and~B, respectively. Although the initial annealing power is comparable across all devices, the final annealing power for Devices~A2 and~B was an order of magnitude higher than that for Device~A1.

This large difference in annealing power resulted in distinct $I$–$V_\text{B}$ behaviors after the initial (black) and final (red) anneals, as shown in Figs.~\ref{Fig. S3}(d)–(f). While Device~A1 remained a closed quantum dot (high resistance and a clear blockade region), Devices~A2 and~B evolved from closed to open quantum dots (high conductance and minimal blockade). Annealing is also known to reduce the oxygen content of Au films and modify their work function~\cite{Chaves14, Heinze02}. This modifies the amount of charge transferred to the SWCNT sections (source/drain) beneath the Au film.

In the main text, we showed that the SWCNT channels in our devices are n-doped (electron-dominated). For low-power annealing (Device~A1), the SWCNT contacts remain p-doped, forming thick tunnel barriers between the channel and contacts, resulting in a closed quantum dot. For high-power annealing (Devices~A2 and~B), the contact regions switch from p-type to n-type, producing weak tunnel barriers and open quantum dots. This behavior is consistent with our previous observation of p-doped graphene contacts converting to n-type after high-power annealing~\cite{McRae24}, and with the known electron–hole asymmetry we reported earlier in SWCNT transistors~\cite{McRae17}.

\subsection{Mechanical calibration and reproducibility}
The mechanical strain range was calibrated by first locating the pushing screw displacement point $\Delta z$ (as shown in Fig. 1 of the main text) where each SWCNT began to stretch. We note that, even when the supporting substrate is slightly bent, a nanotube may not yet be under strain due to some initial built-in slack.

\begin{figure}[!htbp]
\centering
\includegraphics[scale=1.2]{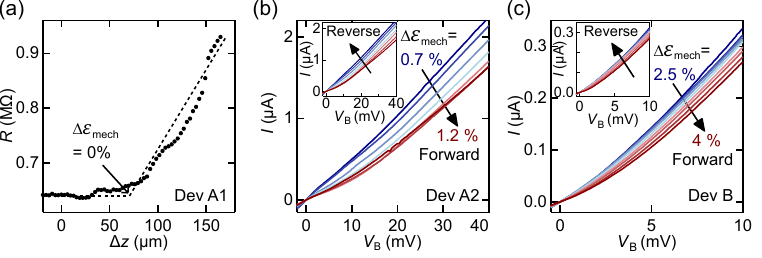}
\caption []{Mechanical calibration and reproducibility. (a) Resistance of Device A1 at $V_\text{B}=135$~mV and $V_\text{G}=5$~V as a function of $\Delta z$. The intersection of the low and high slope fits to the data defined the zero strain position ($\Delta \varepsilon_\text{mech} = 0\%$). The data were reproducible and reversible (see Fig.\ 4(d) of the main text). (b) $I$–$V_\text{B}$ characteristics for Device~A2 as $\Delta \varepsilon_\text{mech}$ varied from 0.7 $\%$ to 1.2 $\%$. Forward (main) and reverse (inset) sweeps showed identical behavior, confirming reproducible (elastic) mechanical strain. (c) Forward and reverse strain sweeps in Device~B.}
\label{Fig. S4}
\end{figure}

We found that the displacement $\Delta z$ at which minor, and fluctuating, resistance changes first appeared was the same in all devices. This onset corresponded to the point where the pushing screw started to bend the substrate and moved the gold clamps. This position defined the substrate’s zero position ($\Delta z = 0$). Because SWCNT channels can have variable mechanical slack upon fabrication, each device had, beyond the substrate's zero position, its own zero-strain point ($\Delta \varepsilon_\text{mech} =$ 0 $\%$) where major (and fully reversible) resistance changes started. We determined the location of these zero strain points by fitting the resistance $R$ versus $\Delta z$ as shown in Fig.~\ref{Fig. S4}(a) for Device A1.

After determining $\Delta \varepsilon_\text{mech} =$ 0 $\%$ for Device A1, we calculated the applied mechanical strain using the device geometry\cite{Champagne05}, $\Delta\varepsilon_{\text{mech}} = \Delta x/L = [(3L_\text{sus}t/D^{2})\Delta z] /L$, where $L_\text{sus}$ is the total suspension length, $L$ is the channel length, $t$ is the chip thickness and $D$ is the distance between two substrate anchoring points (brown rectangles in Fig. 1(a) of main text). Figure~\ref{Fig. S4}(b) shows the $I$–$V_\text{B}$ curves for Device~A2 over the range $\Delta\varepsilon_\text{mech}=$ 0.7 $\%$ – 1.2 $\%$, and the inset presents the data for the reverse mechanical sweep. The forward and reverse sweeps exhibited identical behavior. Similarly reproducible data for Devices~A2 and~B are shown in Fig.~1(f) of the main text and Fig.~\ref{Fig. S4}(c), respectively. These data confirmed the reproducibility and reversibility of the mechanical strain in all three devices.

\section{Coulomb diamond shifts in Device A1}
\label{SM_Fig2}
In Figs.\ 2(a)-(b) of the main text, we showed the Coulomb diamonds for Device~A1 at $\Delta\varepsilon_\text{mech}=$ 0 $\%$ and 0.4 $\%$, respectively, to demonstrate the strain-induced $V_{\text{G}}$ shift. In this section, we present a more detailed analysis of these diamonds and additional data at other strain values to illustrate the continuous strain-induced $V_{\text{G}}$ shift.

\begin{figure}[!htbp]
\centering
\includegraphics[scale=1.2]{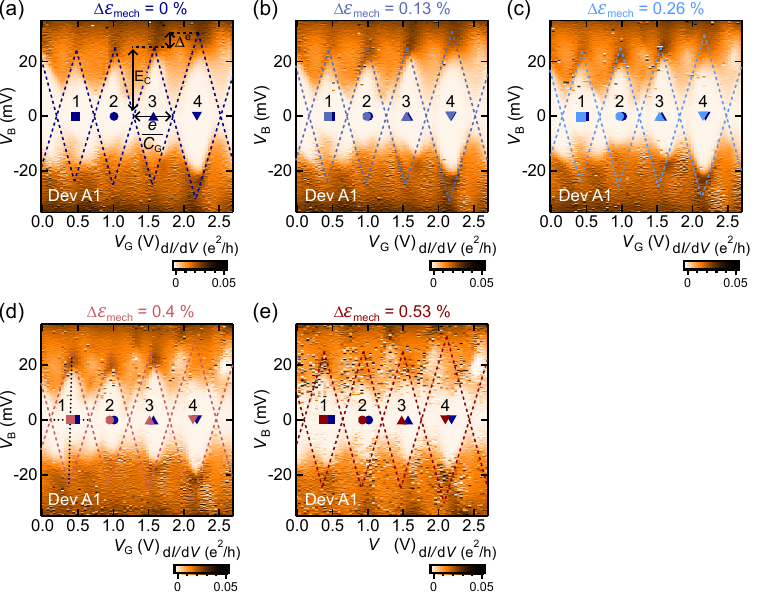}
\caption []{Shifts of the Coulomb diamonds in Device A1 under mechanical strain. (a) At $\Delta\varepsilon_\text{mech}=$ 0 $\%$, Device A1 shows well-defined Coulomb diamonds. The charging energy $E_\text{C}=25$~meV is obtained from the $V_{\text{B}}$ height of the diamond. The width of the diamond depends on the gate capacitance as $e/C_\text{G}$. The $V_{\text{B}}$ height difference between diamonds 3 and 4 is $\approx\Delta^e\approx6$~meV and is the energy level spacing between the subbands. Each diamond centers (1,2,3,4) is marked with a symbol of a different shape. (b)–(e) $dI/dV$-$V_\text{B}$-$V_\text{G}$ in Device A1 at $\Delta\varepsilon_\text{mech}=$ 0.13 $\%$, 0.26 $\%$, 0.4 $\%$, and 0.53 $\%$. The dark blue markers show the original (0 $\%$) position of the diamonds, while the other color shows the position of the diamonds at each strain value. All diamonds shift monotonically leftward with increasing strain.}
\label{Fig. S5}
\end{figure}

Figure~\ref{Fig. S5} shows the $dI/dV$–$V_\text{B}$–$V_\text{G}$ data for Device~A1 at $\Delta\varepsilon_\text{mech}=$ 0 $\%$, 0.13 $\%$, 0.25 $\%$, 0.4 $\%$, and 0.53 $\%$. Diamonds~1–3 have similar widths of approximately 0.56~V, from which the gate capacitance is extracted as $C_\text{G}=e/0.56$ V $=$ 0.29~aF, where $e$ is the elementary charge. The channel length of Device A1 can then be estimated using a wire-over-plane capacitor model\cite{McRae17,Island12}:

\begin{equation}
\label{Eq:S1}
\frac{C_\text{G}}{L_\text{G}} =
\frac{2\pi \varepsilon_\text{ox}}
{\dfrac{\varepsilon_\text{ox}}{\varepsilon_\text{vac}}
\cosh^{-1}\!\left(\frac{t_\text{vac}}{r}\right)
+ \cosh^{-1}\!\left(\frac{t_\text{vac} + t_\text{ox}}{r + t_\text{vac}}\right)}
\end{equation}

\noindent
where $\varepsilon_\text{ox}$ and $\varepsilon_\text{vac}$ are the permittivities, $t_\text{ox}$ and $t_\text{vac}$ are the thicknesses of the oxide and vacuum spacers, respectively, and $r$ is the nanotube radius ($d/2$). The extracted channel length using Eq. \ref{Eq:S1} is $L=$ 31~nm, which agrees quite well with the measured length of 28~nm in the SEM image of Device A1 shown in Fig.~1 of the main text.

Diamonds~1–3 have similar heights of approximately 25~mV, corresponding to the charging energy $E_\text{C}$ of the QD in Device A1. Diamond~4 has a larger height, corresponding to the sum of $E_\text{C}$ and the energy level spacing between the subbands of the SWCNT (4-fold degenerate), $\Delta^{e}\approx6$~meV. A similar $\Delta^{e}$ was observed in suspended SWCNT quantum dots with similar channel lengths in previous work~\cite{McRae17}.

The centers of the Coulomb diamonds labelled 1,2,3,4 in Fig. \ref{Fig. S5} are marked with squares, circle, upward-triangle, and downward-triangle symbols, respectively. As strain is applied to Device A1, a continuous leftward $V_{\text{G}}$ shift of all diamonds was observed. We extracted the $\Delta V_{\text{G}}$ shifts of all four diamonds, at each $\Delta \varepsilon_{\text{mech}}$ value and plotted them in Fig.~2 of the main text. Because the diamond widths are also proportional to the QD charging energy, a conversion factor between $V_{\text{B}}$ and $V_{\text{G}}$ is given as $\alpha = E_\text{C}/\text{width} =$ 0.045. This factor was used to calculate the Fermi energy shift $\Delta\mu_{\text{G}}$ corresponding to each $\Delta V_{\text{G}}$, as shown on the right-hand axis in Fig.~2 of the main text.

\section{Further details on the mechanical gating effect in Devices A2 and B}
\label{SM_Fig3}
In this section, we present additional data and analysis of the mechanical gating effect in Devices A2 and B. These devices have weak electrostatic gating ($V_{\text{G}}$ dependence), most likely due to screening of the QD by the gold clamps after the electromigration process. We showed that the mechanical gating effect is still effective in these devices, and provides an alternate way to gate very small scale molecular devices.

\begin{figure}[!htbp]
\centering
\includegraphics[scale=1.2]{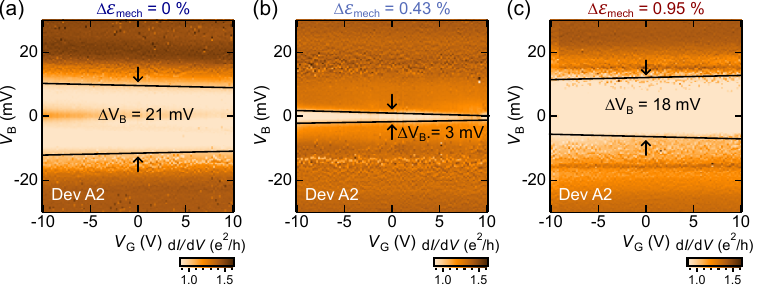}
\caption []{Left shifts of a $dI/dV_{\text{B}-V_{\text{B}-V_{\text{G}}}}$ Coulomb diamond with increasing strain in Device~A2. (a) At $\Delta\varepsilon_\text{mech}=$ 0 $\%$, we see the right-hand side of a Coulomb diamond (N) whose height at position $V_\text{G}=0$~V is 21~mV. (b) At $\Delta\varepsilon_\text{mech}=$ 0.43 $\%$, the diamond shifts left and narrows its $V_\text{G}=$ 0 height to 3~mV. (c) At $\Delta\varepsilon_\text{mech}=$0.95 $\%$, the $N$th diamond has shifted further left (out of $V_{\text{G}}$ range) such that we see the left-hand side of the $N+1$ diamond with a $V_\text{G}=0$ height of 18~mV.}
\label{Fig. S6}
\end{figure}

Figure~\ref{Fig. S6} presents $dI/dV$–$V_\text{B}$–$V_\text{G}$ data for Device~A2 at $\Delta\varepsilon_\text{mech}=$ 0 $\%$, 0.43 $\%$, and 0.95 $\%$. A clear leftward shift of the Coulomb diamond with increasing strain was observed, similar to the one for Device~A1. The measured $\Delta V_{\text{B}}$ height of the diamond at $V_\text{G}=0$~V decreases from 21~mV (Fig. \ref{Fig. S6}(a)) to 3~mV (Fig. \ref{Fig. S6}(b)) as the diamond shifts to the left. When increasing the strain further the Nth diamond shift out of $V_{\text{G}}$ range to the left and we begin to see the following diamond (N+1) opening up with $\Delta V_{\text{B}}=$ 18~mV (Fig.\ref{Fig. S6}(c)). We emphasize that data is available at all intermediate values of $\Delta\varepsilon_\text{mech}$ and shows a smooth an linear $V_{\text{G}}$ shift to the left. The full data set showing $\Delta V_{\text{B}}$ vs $\Delta\varepsilon_\text{mech}$ extracted at $V_\text{G}=0$~V for Device A2 is shown in the main Fig. 3(b).

From the same data set, we extracted the value of the gate shifts $\Delta V_\text{G}$ at each strain by extrapolating the diamond-edge slopes to find the $V_{\text{G}}$ diamond intersection points. We then obtained the Fermi energy shifts using, as previously, $\Delta\mu_\text{G} = \alpha e \Delta V_\text{G}$, where $\alpha = C_\text{G}/(C_\text{S}+C_\text{G}+C_\text{D}) =$ 3 $\times10^{-5}$. The $\Delta\mu_\text{G}$ vs $\Delta\varepsilon_\text{mech}$ for Device A2 are shown in main Fig. 3(c).

\begin{figure}[!htbp]
\centering
\includegraphics[scale=1.2]{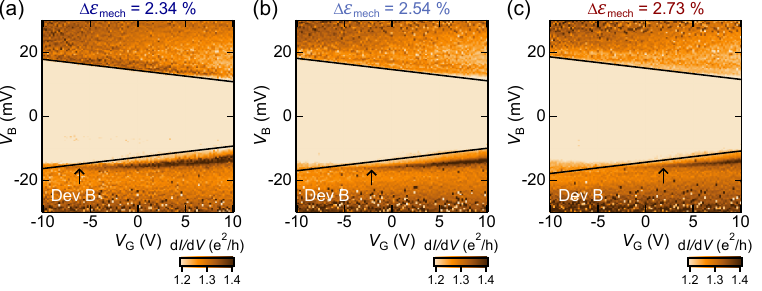}
\caption []{Strain-induced Coulomb diamond shifts in Device~B. (a)–(c) $dI/dV$–$V_\text{B}$–$V_\text{G}$ maps at $\Delta\varepsilon_\text{mech}=$ 2.34 $\%$, 2.54 $\%$, and 2.73 $\%$, respectively. The entire data set shifts rightward, linearly and reproducibly, with increasing strain.}
\label{Fig. S7}
\end{figure}

In contrast to Device~A1 and~A2, the $dI/dV_{\text{B}}$–$V_\text{B}$–$V_\text{G}$ data for Device~B shows a rightward $V_{\text{G}}$ shift of the Coulomb diamonds with increasing strain, as shown in Fig.~\ref{Fig. S7}. The shifts can be observed most easily by tracking the $\Delta V_{\text{G}}$ positions of the $dI/dV_{\text{B}}$ feature indicated by the black arrow. The right shift direction confirms that the $\Delta V_{\text{G}}$ and its underlying Fermi energy shifts with strain, have an additional contribution (mechanism) beyond the strain-induced scalar potential (chirality-independent) which create leftward shifts. This second contribution is from the chirality-dependent vector potentials, also arising from strain, that modify the band structure and open the band gap in the SWCNT. This strain-induced vector potential can shift the Fermi energy in either direction. The quantitative details of the underlying mechanisms of the observed mechanical gating will be discussed in the following section.

\section{Additional information on physical origin of the mechanical gating}
\label{SM_Fig4}
Figure~4 of the main text and the related discussion demonstrate that the mechanical gating observed in our experiments originates from strain-dependent SWCNT band structure modifications rather than from QD capacitance changes. In this section, we present additional data and analysis confirming that our devices are fully clamped mechanically and exhibit no significant capacitance change over a broad strain range. We then describe the theoretical model used to reproduce the transport behavior of our SWCNT devices. In addition to the theoretical calculation of the transport spectrum in Device A shown in Fig.~4(g), we show below the theoretical calculation results for Device~B.

\subsection{Evidence of full clamping: constant capacitance and absence of mechanical hysteresis}
For a free-end quantum dot (Fig.~4(a) of the main text), the distance between the QD and one of the gold clamps would change with mechanical motion, altering the source-QD ($C_{\text{S}}$) or drain-QD ($C_{\text{D}}$) capacitances and thus modifying the Coulomb diamond shapes. To rule out this effect, we show below that the capacitances remain constant over a broad range of mechanical motion. We then present highly reproducible transport data obtained during both forward and reverse mechanical sweeps, providing further evidence that the nanotubes are firmly clamped.

\begin{figure}[!htbp]
\centering
\includegraphics[scale=1.2]{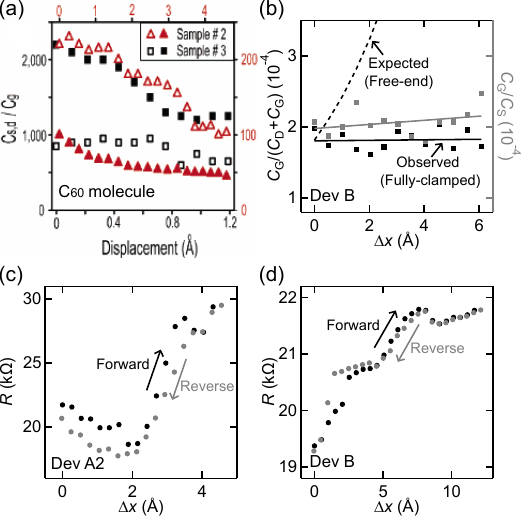}
\caption []{Evidence of full clamping in our SWCNT devices. (a) Capacitance variation in previously reported C$_{60}$-QDs as their gold clamps moved apart. This Figure was adapted from Ref.~\cite{Champagne05}. (b) The capacitance ratios predicted by the free-end QD model (dashed line) shows much larger variations than observed experimentally (black and grey solid markers). (c)–(d) $R$–$\Delta x$ data for Devices~A2 and~B, during forward (black markers) and reverse (grey markers) mechanical sweeps.}
\label{Fig. S8}
\end{figure}

Figure ~\ref{Fig. S8}(a) was reproduced from previous work\cite{Champagne05} and shows the capacitance change measured in two C$_{60}$ quantum dot transistors (square and triangular markers) that were fabricated with the same geometry and nanogap fabrication methods as in our devices (i.e. electromigration of suspended gold breakjunctions). We stress however that the mechanical anchoring of the QDs was vastly different in our SWCNT devices compared to the previously reported C$_{60}$ transistors. We purposefully buried long SWCNT sections ($\geq$ a few microns) underneath each gold clamp to anchor mechanically the QD. In contrast, the C$_{60}$ molecules were dissolved in a solvent and simply cast and dried on top of the gold surface before electromigration.

Figure \ref{Fig. S8}(a) shows the Coulomb diamond slopes measured in the C$_{60}$ experiment versus the mechanical displacement, $\Delta x$, of the gold clamps. The solid and open markers correspond to the capacitance ratios $C_\text{D}/C_\text{G}$ and $C_\text{S}/C_\text{G}$, respectively, as measured in two devices. The authors found that the C$_{60}$ QDs moved relative to either their source or drain electrode as $\Delta x$ increased. The corresponding capacitance ratio varied significantly\cite{Champagne05}. Based on these two samples, we estimated that $C_\text{S}$ or $C_\text{D}$ changed by approximately 12 $\%/\AA$ and 42 $\%/$\AA.

To model how this mechanically-tunable QD capacitance effect would modify our own SWCNT-QD transport data, we used an intermediate value of 25 $\% / \AA$ and calculated $C_\text{D} = C_{\text{D,0}}(1 - 0.25)^{\Delta x/\text{\AA}}$, where $C_{\text{D,0}}$ denotes the initial capacitance. The resulting calculated ratio, $C_\text{G}/(C_\text{D}+C_\text{G})$, is plotted versus $\Delta x$ as the dashed line in Fig. ~\ref{Fig. S8}(b). This free-end QD behavior is to be compared with our measured QD capacitance ratios in Device B, as shown by the black markers for $C_\text{G}/(C_\text{D}+C_\text{G})$ and by the grey markers for $C_\text{G}/C_\text{S}$ versus the clamp displacement $\Delta x$. Both of the measured capacitance ratios remain nearly constant  across displacement range of 6~\AA\, and the solid black and gray lines are fits. In addition, Figure 4(c) in the main text shows similar measured data versus a free-end calculation for Device A2.

For all three of our measured Devices, we do not see any significant change in the capacitance ratios versus $\Delta x$, while the free-end QD model (Fig. 4(a)) predicts much larger changes. These results rule out QD capacitance variations as the dominant origin of our observed QD mechanical gating. Additional evidence that our QDs are firmly clamped at both ends (source and drain) can be seen in Figs.~4(d) and \ref{Fig. S8}(c)–(d)). These data from our three Devices (forward mechanical sweep in black and reverse sweep in gray) show that in all of our devices the mechanical dependence of transport is fully reversible over the studied mechanical range. This behavior was reproducible over multiple mechanical sweeps. This confirms that the QD channels were fully clamped (no slippage) and strained elastically (no hysteresis).

\subsection{An applied theory of QTS in quasi-metallic SWCNTs}
In a fully clamped suspended SWCNT transistor (Fig. 4(e)), mechanical motion of the gold clamps induces a uniaxial and uniform strain in the channel. This modifies the SWCNT band structure\cite{Huang25,McRae19,Amorim16,CastroNeto09,Kitt12} through scalar, $\phi_{\varepsilon}$, and vector, $\boldsymbol{A}$, potentials as illustrated in Fig.~4(f) of the main text. In this subsection, we summarize how these potentials affect ballistic transport in SWCNTs, as reported in detail in previous theoretical work\cite{Huang25}. We then discuss the quantitative connection between the theoretical ballistic SWCNT transport behavior and our SWCNT-QD measurements.

The Hamiltonian of charge carriers (charge $e$) in the strained SWCNT channel, incorporating the scalar and vector potentials, and unstrained SWCNT contacts are written as Eqs.~\ref{Eq:H_channel} and~\ref{Eq:H_contacts}, respectively\cite{Huang25, McRae19,Pellegrino11,Naumis17,Tworzydlo06}:

\begin{equation} \label{Eq:H_channel}
H_{K_{i},\text{channel}} = \hbar v_{F}  \boldsymbol{\sigma} \cdot (\bar{I} + (1 - \beta)\bar{\boldsymbol{\varepsilon}}) \cdot \boldsymbol{\tilde{k}} + e\phi_{\varepsilon} + \Delta\mu_{\text{G}}
\end{equation}

\begin{equation} \label{Eq:H_contacts}
H_{K_{i},\text{contact}} = \hbar v_{F} \boldsymbol{\sigma} \cdot \boldsymbol{k} + \mu_{\text{contact}}
\end{equation}

where $K_i$ denotes the valley, and $\boldsymbol{\tilde{k}} = \boldsymbol{k} - \boldsymbol{A_i}$ and $\boldsymbol{k}$ are the electron wavevectors in the SWCNT channel and contacts, respectively. The Fermi velocity is $v_F = 8.8 \times 10^{5}$~m/s, $\bar{I}$ is the identity matrix, and $\bar{\boldsymbol{\varepsilon}}$ is the strain tensor. The electronic Gr\"uneisen parameter is $\beta \approx 2.5$~\cite{Naumis17}. The factor $\bar{I} + (1 - \beta)\bar{\boldsymbol{\varepsilon}}$ represents the strain-dependent correction to the Fermi velocity. The pseudospin operator $\boldsymbol{\sigma}$ corresponds to the 2D Pauli matrix vector, $\phi_{\varepsilon}$ is the strain-induced scalar potential, $\Delta\mu_{\text{G}}$ is the gate-induced electrostatic energy, and $\mu_{\text{contact}} =$ 0 to 0.25 eV is the Fermi level in the source and drain SWCNT contact regions due to the charge transfer from the physisorbed gold~\cite{Naumis17}.

We note that the theoretical calculations are insensitive to changes in $\mu_{\text{contact}}$ over the entire range of plausible values (0 to 0.25 eV) because in all cases the SWCNT channel supports a single conduction mode due to its subband spacing. In the device’s $x$–$y$ coordinates, $x$ is aligned along the tube axis and $y$ along the circumference. The strain tensor $\bar{\boldsymbol{\varepsilon}}$ components are then $\varepsilon_{xx} = \varepsilon$, $\varepsilon_{yy} = -\nu\varepsilon$, and $\varepsilon_{xy} = \varepsilon_{yx} = 0$, with Poisson ratio $\nu = 0.165$~\cite{Naumis17}.

The vector potentials $\boldsymbol{A_i}$ are the sum of two contributions, $\boldsymbol{A_{i,\text{lat}}}$ and $\boldsymbol{A_{\text{hop}}}$ which originate from modifications in the nearest-neighbor atomic positions and hopping amplitudes, respectively~\cite{Kitt12}. Our previous theoretical work\cite{Huang25} showed that the $\boldsymbol{A_{i,\text{lat}}}$ terms are canceled by the subband momentum shifts induced by uniaxial strain in the SWCNT (because the circumference of the tube shrinks as it is strained). This leaves only the hopping term as relevant, and it is expressed in Eq.~\ref{Eq:A}:

\begin{equation} \label{Eq:A}
\mathbf{A}_{\text{hop}} = \frac{\beta \varepsilon (1 + \nu)}{2a}
\begin{pmatrix}
\sin 3\theta_h \\
\cos 3\theta_h
\end{pmatrix}
\end{equation}
\noindent
where $a = 1.42$~\text{\AA} is the unstrained nearest-neighbor carbon–carbon bond length, and $\theta_h$ is the chiral angle given by $\theta_h = \tan^{-1}(\sqrt{3}m/(2n + m))$.

\begin{figure}[!htbp]
\centering
\includegraphics[scale=1.2]{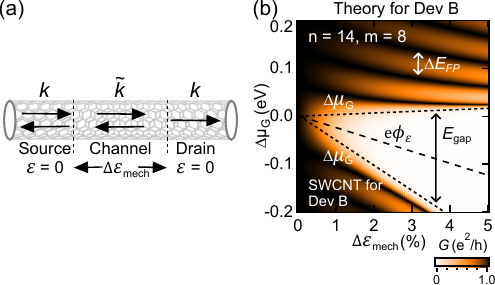}
\caption []{Applied-theory calculations for quantum transport in strained SWCNTs\cite{Huang25}. (a) Diagram showing a 1D representation of the reflection and transmission of charge carriers across a strained SWCNT channel and unstrained SWCNT source/drain contacts. (b) Calculated conductance $G$ as a function of $\Delta\mu_\text{G}$ and $\varepsilon_\text{mech}$ for the reported Device~B using this applied theory.}
\label{Fig. S9}
\end{figure}

We solved for the transmission amplitude $t_{\xi}$ to calculate the conductance $G$ across the transistor, using the longitudinal boundary conditions at the source–channel and channel–drain interfaces (Fig.~\ref{Fig. S9}(a)) for each valley $\xi = \pm 1$ (corresponding to $K$ and $K'$). The transmission probability is $T{_\xi} = |t_{\xi}|^{2}$, as given by Eq.~\ref{Eq:T_n_Str_A}:

\begin{equation} \label{Eq:T_n_Str_A}
T_{\xi} = \frac{(k_x \tilde{k}_x v_{xx})^2}
{(k_x \tilde{k}_x v_{xx})^2 \cos^2(\tilde{k}_x L) +
(k_F \tilde{k}_F - k_y (k_y - \xi A_{\text{hop},y}) v_{yy})^2 \sin^2(\tilde{k}_x L)}
\end{equation}
\noindent
where $k_F = \mu_{\text{contact}} / (\hbar v_F)$ and $\tilde{k}_F = \mu_{\text{channel}} / (\hbar v_F)$. The strain-modified velocity factors are $v_{xx} = 1 + (1 - \beta)\varepsilon_{\text{mech}}$ and $v_{yy} = 1 - (1 - \beta)\nu\varepsilon_{\text{mech}}$. Therefore,
$k_x = \sqrt{k_F^2 - k_y^2}$ and
$\tilde{k}_x = v_{xx}^{-1} \sqrt{\tilde{k}_F^2 - (v_{yy} (k_y - \xi A_{\text{hop},y}))^2}$.

Finally, we can obtain the ballistic conductance using

\begin{equation} \label{Eq:G}
G = \frac{4e^2}{h} \frac{1}{2} \sum_{\xi} T_{\xi}
\end{equation}
\noindent

We computed the $G$–$\Delta\mu_\text{G}$–$\varepsilon_\text{mech}$ characteristics shown in Fig.~4(g) of the main text and Fig.~\ref{Fig. S9}(b) for quasi-metallic nanotubes (22,16), matching Device A, and (14,8), matching Device B, under strain $\varepsilon_{\text{mech}}$ varying from 0 to 5$\%$. A bandgap opens up with increasing strain due to the vector potential, while the spectral (Fabry-P\'{e}rot) interferences shift as a result of both the scalar and vector potentials. These effects define the upper and lower bandgap edges (short dashed lines in Fig.~\ref{Fig. S9}(b)), corresponding to the $\Delta \mu_{\text{G}}$ threshold where the current turns on as a function of strain.

The sign and magnitude of this strain-induced $\Delta \mu_{\text{G}}$ shift depend on the nanotube chirality and carrier type. This explains the opposite shift direction in Devices~A1/A2 (on the same tube) compared to Device~B. Using the measured values of this chirality-dependent $\Delta \mu_{\text{G}}$ shift, as well as the measured tube diameters $d = a\sqrt{n^{2} + m^{2} + nm}/\pi$\cite{Laird15}, we narrowed down the only possible chiralities in our devices to $(n,m) = (22,16)$ or $(21,15)$ for Devices~A1/A2 and to $(14,8)$ for Device~B (see Fig.~4(h) of the main text). The slope of $\Delta \mu_{\text{G}}$ vs $\Delta \varepsilon_{\text{mech}}$ varies rapidly even with minor chirality changes as can be seen in Fig. 4(h), and from Eq. S7 below.

\begin{equation}
\Delta\mu_\text{G} = -g_{\varepsilon}(1-\nu)\varepsilon_{\text{mech}} \pm \hbar v_{\text{F}}\frac{\beta\varepsilon(1+\nu)}{2a}\cos[3(\tan^{-1}\!\left(\frac{\sqrt{3}\, m}{2n + m}\right))]
\end{equation}

Therefore, when combined with the tube diameter measurements, the $\Delta \mu_{\text{G}}$ vs $\Delta \varepsilon_{\text{mech}}$ data enable a precise chirality measurement in our devices based on their charge transport.

\bibliographystyle{api2}
\linespread{1.0}